\documentstyle[12pt]{article}
\begin{document}

\vspace*{1cm}
\begin{center}
{ \Huge\bf
Restoration of the Past and Three Principles of Time}

\vspace*{1cm}

{\large

Alexandr K.\ Guts}
\vspace*{1cm}

{\small

Department of Mathematics, Omsk State University \\
644077 Omsk-77 RUSSIA
\\
\vspace*{0.5cm}
E-mail: guts@univer.omsk.su  \\
\vspace*{0.5cm}
March 5, 1997\\    }
\vspace{.5in}
ABSTRACT
\end{center}
{\small

Can the Past be restored? Poincar\'e and Costa de Beauregard showed that
the past is not restored statistically. This follows from Bayes formula.
It is  mith that history can make this and moreover that history is
created for this. Historian is sure that he is free for such work.
In this mithological world one are living not only historians but
all people. The fear of death is a cause of such confidence.
In this note three Principles of Time will be formulated which
say that historians can not give us the truth text-book of History
of any society.

}


\vspace*{2cm}

Can a reseacher restore the events of the past epoches without distortion?
It is  mith that history can make this and moreover that history is
created for this. Historian is sure that he is free for such work.
In this mithological world one are living not only historians but
all people. The fear of death is a cause of such confidence.
In this note three Principles of Time will be formulated which
say that historians can not give us the truth text-book of History
of any society.

\section{ Bayes Principle}          

In the book \cite{Tim}, Costa de Beauregard formulates
the following

{\bf Bayes Principle}. {\it Restoration of the Past can be made only
in the case when the a priori's probabilities are known, i.e. if some
knowledge about the Past is assumed the knowledge of the Present
can only specify one}.

In other words, although the Future is statistically predicted the Past
is not statistically restored. In  \cite{Tim}, Costa de Beauregard  gives
very instructive example which belongs to Poincar\'e.

The a priori's probabilities are determinated on the basis historical
documents, therefore there exists the hope that in future the document
will be found which will ascertain the truth. But this hope is vain as
it follows from the third Principle of time (see \S 3).

\section{Principle of the uncertainty of description}  

This  principle has the following form
$$
\Delta D \Delta t \geq c_1                       \eqno(1)
$$
where $\Delta D$ is historical uncertainty, i.e. the number of
contradictory details in description of historical event which
occured for time $\Delta t$, and $c_1 > 0$ is some constant.

In other words, the less time of life of investigated historical event the
more contradictory details.

But it is known that many crimes are successfully investigated. Does it
mean that second Principle is false? No. There exists the concept
"time of  prescription". It is required the Third Principle of Time
(see \S 3).

\section{Principle of the interaction of epoches} 

Principle of the uncertainty of historical description acts only under
condition of realization another Principle which is called the
Principle of the interaction of epoches and which asserts that
historical uncertainty $\Delta D$ is more when  the investigated epoch
lies farther from present epoch:
$$
	 \Delta D \leq c_2 \Delta\tau,             \eqno(2)
$$
where $\Delta\tau$ is time interval between present and
investigated epoches, $c_2 >0$ is some constant.
In other words, the more antique epoch the less chance to ascertain the
truth.

The formula (2) is similar to the Shenon Principle which asserts that
precision of information (from one system to different one)
which can be transfered and received is proportional to the time
that runs out
$$
 \Delta I \leq k \Delta\tau .
$$
It follows  from (1) and (2) that
$$
\Delta\tau\Delta t \geq  c_1c_2^{-1}
$$
or
$$
\Delta t \geq  c_1c_2^{-1}\frac{1}{\Delta\tau} \to 0 \quad under \quad
\Delta\tau\to \infty               \eqno(3)
$$
It means that for restoration of the events which belong to time
interval $\Delta t$ of interesting for us epoch must not be too near
to the present one.
The constants $c_1$ and $c_2$ are determinated from (1)-(3).
 They must have such values to be possible the investigations of crimes of
the recent past.
For example,  it must exist the possibility to restore uniquely
all events, i.e. $\Delta D =0$ if a crime was done $1 \ hour$ ago,
$\Delta \tau = 60 \ min$.
In other words,
$c_1 \leq 1 \ [min^{-1}]$, and $c_2\leq 1/60 \ [min^{-1}] $. One can take
$c_1 =1 \ [min^{-1}]$, and $c_2= 1/60 \ [min^{-1}] $.

We see that the problem of determination of constants $c_1$ and $c_2$
is not easy.

One can suppose that interaction of epoches has more composite
oscillating character with increasing amplitude
$$
\Delta D\leq c_2 \Delta\tau f(\Delta\tau)
\cos\left(\frac{2\pi\Delta\tau}{T} \right),
 \eqno(4)
$$
where $T$ is period and $f(\Delta\tau)\geq 0$ is non-decreasing
function.  The negative value of right side of (4) on segment
$\Delta\tau\in[T/4+nT, 3T/4+nT]$ must be interpreted as epoches
for which one-valued restoration of events is possible.

Uncertainty of historical description insreases for epoches
$\Delta\tau = nT$ when we deepen in the Past.

  Principle (4) is held for the resilient space-time $V^4$,
i.e. $V^4$ when is a resilient leaf of some foliation ${\cal F}$ in
five-dimensional space.    The resilient
space-time winds round itself. A movement along 5-th coordinate
gives the infinite piercing
of space-time $V^4$  at the points of Past and Future
\cite{Gu1, Gu2, Gu3, Gu4}.  The past epoches winds round present epoch
and become nearer  and nearer  as this see the observer living in
five-dimensional world. The (almost) oscillation follows from this.
Let that the coming has value of the Plank's length
$L \sim 10^{-33} cm$ for past epoches $\Delta\tau = nT, n > n_0$.
Then natural fluctuations of 5-metric (electro-gravity-scalar field)
will make foam topology of four-dimensional space-time. Hence the
present epoch will connect with the past epoches $\Delta\tau = nT, n > n_0$
by means of four-dimensional wormholes.
Hence  the events of present and past
are not differed under quantum point of view.
The Past is mixed with the Present.
Such intermixing of Past with Present must have some projection
on the events of macro-world. It follows from \cite{Gu1, Gu2} where
it is shown that four-dimensional macro-wormholes connecting Past and
Present  can appear under perturbations of scalar fields.

Maybe oscillating character of  intermixing of Past with Present was
exposed in the statistical investigations of historical events
by A.T.Fomenko
\cite{Fo}.

\end{document}